\documentclass[12pt,preprint]{aastex}
\usepackage{graphicx}
\usepackage{amsmath}
\usepackage{amssymb}
\usepackage{bm}
\usepackage{natbib}
\def\vn{\mathbf{n}}
\newcommand{\R}{\mathbb{R}}
\def\cR{\mathcal{R}}
\def\bra{\langle}
\def\ket{\rangle}
\def\mean#1{\bra {#1} \ket}
\begin{document}
\title{On discrete symmetries and relic radiation anisotropy}
\author{M.V.Altaisky\thanks{also at Joint Institute for Nuclear Research, Dubna, 141980, Russia}}
\affil{Space Research Institute RAS, Profsoyuznaya 84/32, Moscow, 117997, 
Russia}
\email{altaisky@mx.iki.rssi.ru}
\and
\author{N.E.Kaputkina}
\affil{National  University of Science and Technology ``MISiS'', Leninsky prospect 4, Moscow, 119049, Russia}
\email{nataly@misis.ru}
\date{Revised: Nov 15, 2011}
\begin{abstract}
It is argued that wide-angle correlations of the Cosmic Microwave Background Radiation (CMBR) temperature fluctuations measured by Wilkinson Microwave Anisotropy Probe (WMAP) mission 
may have a trace of discrete symmetries of quantum gravity
\end{abstract}
\maketitle
The standard cosmological model, the $\Lambda$CDM, assumes the matter to be forever expanding in utmost flat universe under the pressure of dark 
energy. In early Universe, before the recombination of electrons 
and protons and the formation of galaxies, the electron-proton plasma was assumed homogeneous. The formation of galaxies was considered as a consequence of gravitational instability of small initial density fluctuations $\delta \rho/\rho \sim  10^{-5}$, constrained by the results of WMAP mission \citep{wmap5}.  In post-recombination epoch the gas of neutral atoms became utmost transparent to photons. The 
CMBR, with the present temperature 2.7K, was no longer scattered by the matter (except for the scattering on the free electrons of galaxies \citep{SZ1969s}), and therefore carries  invaluable information on the early Universe.

The $\Lambda$CDM model is an inflationary model, which depends on six parameters: the barion density $\Omega_b h^2$, the cold dark matter density $\Omega_c h^2$, a cosmological constant $\Omega_\Lambda$, the spectral index 
of fluctuations $n_s$, the scalar fluctuation amplitude $\Delta_\cR^2$, and the optical reionization depth $\tau$
\citep{Dunkley2009,Larson2011}. The model assumes a nearly Gaussian spectrum of initial fluctuations with 
statistical isotropy over the sky \citep{PW1967CMBR}.  Any deviation from isotropy would be a challenge for a new physics\citep{Efstathiou2003,Tegmark2003}.

Although up to the recent WMAP data  the $\Lambda$CDM model remains a good fit of the observed microwave sky 
\citep{Komatsu2011}, 
the key feature displayed by WMAP \citep{BennetEtAl2011} and previous experiments \citep{Smoot1992,strukov1993anisotropy,Bennet1996}, is that the Universe is isotropic in the mean, but 
anisotropic in correlations. This means there are no preferable directions, but there are preferable angles 
of correlations. The observation of such anisotropy is contr-intuitive from classical physics point of view, 
but seems quite natural in quantum mechanics, like that of the Einstein-Podolsky-Rosen (EPR) correlations \citep{EPR1935}. The anisotropy 
in angle correlations has been receiving constant attention since discovered \citep{Smoot1992}. WMAP mission itself was designed to use the observed correlations of fluctuations to put narrower constraints on cosmological parameters after those obtained by the previous COsmic Background Explore (COBE) mission \citep{Bennet1992,Bennet2003}. 

Two main types of data are available to constrain the cosmological parameters: {\em (i)} the observed correlations in the distribution of galaxies \citep{PG1975}; and {\em (ii)} the observed correlations of the relic radiation \citep{BennetEtAl2011}.
The main instrument for the analysis of both data type remains the decomposition with respect to the  representations of the $SO(3)$ group of rotations in $\R^3$. 
This means the $n$-point correlation 
function $\mean{{u}^{\beta_1}\ldots{u}^{\beta_n} }$ transforms under the spacial rotations according to the law  
\begin{equation}
\mean{{u'}^{\alpha_1}\ldots{u'}^{\alpha_n} } = 
\Lambda^{\alpha_1}_{\beta_1} \ldots \Lambda^{\alpha_n}_{\beta_n}
\mean{{u}^{\beta_1}\ldots{u}^{\beta_n} } \label{gtl}
\end{equation}
where $\Lambda$ is the matrix of $SO(3)$ rotation in 
appropriate representation. Thus a scalar remains invariant under rotations; the $n$-point correlation function 
of a vector field thus transforms according to the 
direct product $\underbrace{\Lambda \otimes \ldots \otimes \Lambda}_{n\ times}$, etc. 

In the COBE and the WMAP data 
the full sky map was decomposed into a series of spherical harmonics $Y_{lm}(\vn)$, which form irreducible 
representations of $SO(3)$ group in $\R^3$:
\begin{equation}
T(\vn) = \sum_{l=0}^\infty \sum_{m=-l}^l a_{lm}Y_{lm}(\vn), \quad \hbox{with\ }
a_{lm}= \int d\vn T(\vn) \overline{Y_{lm}(\vn)},
\end{equation}
where $\vn$ is the unit direction vector. If the CMBR anisotropy is driven by a Gaussian random process with 
random phases and zero mean, then 
\begin{equation}
\mean{a_{lm}\overline{a_{l'm'}}}= C_l \delta_{ll'}\delta_{mm'},
\end{equation}
where $C_l$ is the angular power spectrum. 
The statistical distribution of the $a_{lm}$ coefficients has been thoroughly studied by the WMAP team, 
with no significant deviation of the gaussianity found \citep{Komatsu2011}. 

At large angles (low multipoles) the main attention has been paid to the quadrupole component, 
which was studied in the framework of different models since first measured by COBE \citep{Bennet1992,Hinshaw1996,Efstathiou2003,de2006cmb,BennetEtAl2011}. 

The pair correlator of two events separated by the angle $\alpha$
\begin{equation}
C(\alpha) =
\langle \Delta T(\theta) \Delta T(\theta+\alpha) \rangle \label{cf}
\end{equation}
is related to the subgroup $SO(2)$ of the rotation group $SO(3)$. The $C(\alpha)$ averages over the orientation  
of the observation plane. Since first measured by COBE the pair correlator \eqref{cf} has 
been parametrized in the form \citep{Wright1992} 
$$
C(\alpha) =  A + B \cos \alpha + C^0_M \exp \left[
-\frac{\alpha^2}{2\sigma^2} \right] ,
$$ 
although the locations of autocorrelation function maxima at 
$0$ and $\frac{2\pi}{3}$ and the minima at $\frac{\pi}{3}$ and $\pi$ suggest another parametrization \citep{ABK1996}:
 \begin{equation}
C(\alpha) =  A + B \cos 3\alpha + C^0_M \exp \left[
-\frac{\alpha^2}{2\sigma^2} \right] \label{cfa3},
\end{equation} 
where $A,B,C^0_m,\sigma$ are constants.

If the correlation function $C(\alpha)$ has 
a maximum around $\alpha = \frac{2\pi}{3}$ and the minimum around  $\alpha = \frac{\pi}{3}$, 
see Fig.5 of \citet{BennetEtAl2011}, the obvious subgroup of $SO(2)$, which can explain such behavior, is 
the group $C_3$ of the rotations $R_k=R_1^k=R\left(\frac{2\pi k}{3}\right), k=0,1,2$: 
\begin{equation}
C_3 = \left\{E, R_1,R_1^2 \right\}. \label{c3def}
\end{equation}
The simplest explanation of the presence of $C_3$ group in the CMBR temperature fluctuations was given in \citet{ABK1996}.
 This explanation is in considering simplicial quantum gravity \citep{AJ1992,AM1992}, 
which enables to describe pure gravity in a consistent way. The solution with coupling to matter fields 
was found for two-dimensional case \citep{BK1990b,BKKM1986}. Starting from Boulatov-Ooguri lattice gravity models
\citep{Boulatov1992,Ooguri1992}, 
which substitute the integral over all possible geometries by a discrete state sum over all 
possible triangulations $Z=\sum_T C_T e^{-S(T)}$, where the action $S(T)$ counts the number of simplexes, 
the theory has merged into spin-foam quantum gravity \citep{BC1998}. The spin foam has a natural interpretation of quantized space-time based on discrete symmetry groups \citep{PR2001,CPR2001}.

The assumption of the gaussianity of 
initial density fluctuations, being purely classical, is not self-sufficient. It leaves the question where from the smooth manifold of Friedmann-Robertson-Walker universe has originated. The formation of a smooth  
manifold by no means follows the Big Bang hypothesis. Instead, the formation of a smooth  
manifold should arise from quantum gravity models. The dynamical triangulation in these models 
allows for a smooth manifold in the limit of the infinite number of simplexes, preserving the initial 
discrete symmetries at a quantum level.
  
A simple toy-model, which assumes the dynamical formation of fractal space with a symmetry group of 
simplex, which covers a $d$-dimensional sphere, has been proposed in \citep{ABK1996}.
The model assumes the dynamical creation of discrete geometry starting from a "point" Universe.  
An alternative model is based on the discrete  
symmetry group of curved dodecahedral Poincare space. In a curved Poincare 
dodecahedral space the spherical dodecahedrons, used as building cells, have the edge angles 
exactly $\frac{2\pi}{3}$, which may explain the correlation maximum observed by WMAP mission 
\citep{LWRLU2003}.

If we base on symmetry only, without any special assumptions on geometry,
we ought to construct $n$-point correlation functions according to the general theory of 
group representations, see e.g. \citet{Hamermesh}.
So the $\Lambda^{\alpha}_{\beta}$ matrices in \eqref{gtl} must the matrices of a given representation 
of symmetry group.
In the simplicial fractal model of \citet{ABK1996} the considered symmetry is the abelian group $C_3$.
The $C_3$ group \eqref{c3def} has three non-equivalent irreducible representations $T^{(1)}$, $T^{(2)}$, $T^{(3)}$, the 
characters of which are shown in Table~\ref{t1} in Appendix.

The representation $ T^{(1)}$ is singlet. It describes the functions not affected by 
$C_3$ transformations. Two other representations $ T^{(2)}$ and $ T^{(3)}$ correspond to the 
functions subjected to the left and the right phase rotations under the $C_3$ transformations.
If the functions $u_i,u_j,u_k,\ldots$ transform according to representations $T^{(i)}$, $T^{(j)}$, $T^{(k)}\ldots$, 
respectively, the $n$-point correlation function  transforms according to the direct product 
$T^{(i)}\otimes T^{(j)} \otimes T^{(k)}\otimes\ldots$. 

The decomposition of the direct product into a sum of irreducible 
representations is casted in the form 
\begin{equation}
T^{(\alpha)}\otimes T^{(\beta)} = \oplus_\gamma m_\gamma T^{(\gamma)} \quad 
\label{dpd}
\end{equation}
with the weights $m_\gamma$ is determined by standard formula \citep{Hamermesh}
\begin{equation}
m_\gamma = \frac{1}{g}\sum_{p\in G} \overline{\chi^\gamma_p} \chi^\alpha_p \chi^\beta_p,
\label{chardec}
\end{equation}
where $g$ is the number of elements in the group $G$, $\chi^\alpha_p$ is the character of the element $p$ in the 
representation $T^{(\alpha)}$. For the group $C_3$ the decomposition \eqref{dpd} has a simply-reducible form 
 \begin{eqnarray}
 \nonumber
 T^{(1)}\otimes T^{(1)} = T^{(1)}, & T^{(1)}\otimes T^{(2)} = T^{(2)}, &  T^{(1)}\otimes T^{(3)} = T^{(3)}, \\
 T^{(2)}\otimes T^{(2)} = T^{(3)}, & T^{(3)}\otimes T^{(3)} = T^{(2)}, & T^{(3)}\otimes T^{(2)} = T^{(1)}. \label{c3mt} 
 \end{eqnarray}
 For definiteness, let us consider the field $u$, which transforms according to the $T^{(2)}$ representation, then 
 \begin{equation}
 \mean{u} \sim T^{(2)},\quad \mean{uu} \sim T^{(3)},\quad \mean{uuu} \sim T^{(1)}.
 \label{mtl}
 \end{equation}
 As it follows from the \eqref{mtl}, starting from the field with nontrivial transformation 
 properties, either $T^{(2)}$ or $T^{(3)}$, the singlet states, i.e., those invariant under any $C_3$ 
 transformation, can be obtained only for triple correlator $\mean{uuu}$, rather than for pair correlator 
 $\mean{uu}$.   

The singletness of the triple correlations in \eqref{mtl} is an interesting fact for the $C_n$ group is the a point subgroup of the $SO(3)$.
The other point subgroups of $SO(3)$ related to polyhedra are nonabelian groups. 
In the simplicial model of \citet{ABK1996} the next to $C_3$ hierarchic level has the symmetry 
group of tetrahedron $T_d$, which includes $C_3$ as a subgroup.
$T_d$ comprises 3 axis of second order ($C_2$) and also ($C_3$) rotations $R_1$ and $R_1^2$ 
for each of four faces.
$T_d$ has 4 irreducible representations (3 one-dimensional representations: $T^{(1)},T^{(2)},T^{(3)}$, and one three-dimensional, $T^{(4)}$, which stand for rotations). Using the character table of $T_d$ group (Table~\ref{t2} in Appendix), we can  see that either of 
representations  $T^{(1)},T^{(2)},T^{(3)}$ yield a singlet in triple correlations.
The application of \eqref{chardec} gives 
the following decomposition table for $T_d$:
\begin{eqnarray}
 \nonumber
 T^{(1)}\otimes T^{(\alpha)} = T^{(\alpha)}, &  T^{(4)}\otimes T^{(\beta)} = T^{(4)}, &
 \alpha=1,2,3,4; \beta=1,2,3, \\
  T^{(2)}\otimes T^{(2)} = T^{(3)}, & \label{t4mt}
 T^{(3)}\otimes T^{(2)} = T^{(1)}, &  T^{(3)}\otimes T^{(3)} = T^{(2)},  \\
 \nonumber
 T^{(4)}\otimes T^{(4)} = T^{(1)}  +& T^{(2)} + T^{(3)} + 2 T^{(4)}. & 
  \end{eqnarray} 
 As it can be seen from multiplication table \eqref{t4mt}, if we take a field $u$ in 
 one-dimensional representation $T^{(2)}$ the relations \eqref{mtl} hold for tetrahedron group. This confirms the conclusions of \citet{ABK1996} that the observed $C_3$ extrema in pair correlations may be manifestations of underline $T_d$ symmetry. 
 
 To test the presence of discrete symmetries, either $C_3$ or $T_d$,  
 we can study the behavior of the three-point correlation functions, specially that with equal angles between the legs. 
 The deviations from singlet, possibly due to polarization, may be important to 
 test the discrete symmetry of the Universe at quantum gravity stage. 
 
 The suggested discrete symmetry may  in some sense be similar to that of quantum chemistry of $NH_3$ and $CH_4$ molecules ($C_3$ and $T_d$ symmetry, respectively). A molecule ''as it is'' before any measurement have no preferable direction, but 
 then measured or participate in chemical reaction, the discrete symmetry is displayed.   

\begin{figure}[ht]
\centering \includegraphics[angle=270,width=0.7\textwidth]{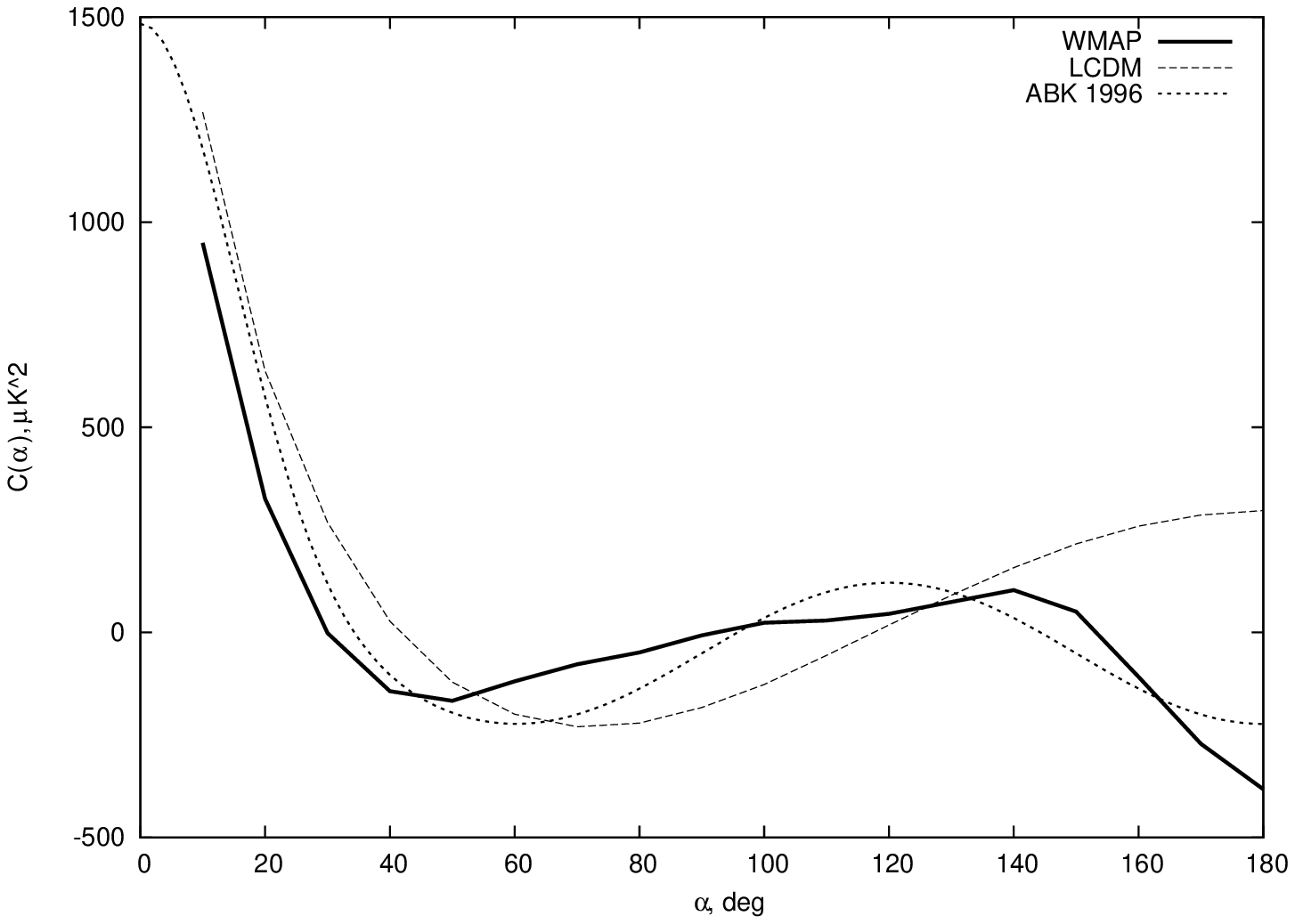}
\caption{Fit of the $C_3$ symmetry assumption (Eq.8  of \citet{ABK1996}) onto WMAP data. The solid line indicates WMAP temperature fluctuations correlation function ; the dashed line 
shows $\Lambda$CDM prediction -- both lines from Fig.5 of \citet{BennetEtAl2011}. By dotted line 
we indicated the prediction of the equation \eqref{cfa3}}
\label{wc:pic}
\end{figure}

To stress the significance of $C_3$ symmetry we fit the recently released 7 year WMAP data 
 (Fig.5 of \citet{BennetEtAl2011}) by the equation \eqref{cfa3}. The result is shown in Fig.~\ref{wc:pic}.
Clearly in the region $\frac{2\pi}{3}\le\alpha\le\pi$ the decreasing behavior of the equation \eqref{cfa3} 
better fits WMAP correlation function then the $\Lambda$CDM curve.

To compare the data fits of the WMAP 7 year data given by $\Lambda CDM$ and the fit given by the model \eqref{cfa3} we have 
calculated pixel to pixel correlation function from WMAP 7 year data 
source from http://lambda.gsfc.nasa.gov, which contains the information on 3145728 pixels, histogramed the data into $n=18$  
bins of size 10 degree each, and compared the goodness of fit of this correlation 
by the equation \eqref{cfa3} and the $\Lambda CDM$ model prediction (given in Fig. 5 of \citet{BennetEtAl2011}) according to the Student $t$-distribution for the variable of $n-1$ degrees of freedom 
\begin{equation}
t_n = \sqrt{n-1}\frac{\bar x}{\bar s},
\end{equation}
where $n$ is the number of bins, $\bar x$ is the estimation of the mean deviation, $\bar s$ is the square root of estimated variance. The results of comparison are summarized in the Table ~\ref{t:tab} below.
\begin{table}[ht]
\begin{center}
\begin{tabular}{|l|rrrrr|}
\hline
       & Number & Bin size & $\bar s$ & $t_n$ & $p$-value \\
       & of bins& degrees  & $\mu K^2$&       &           \\
\hline 
$\Lambda CDM$            & 18     & 10       & 277.3    & -1.76 &  0.096    \\
ABK1996         & 18     & 10       & 120.1    &  1.38 &  0.186    \\   
\hline 
\end{tabular}
\caption{Comparison of the $\Lambda CDM$   and Equation \eqref{cfa3} model fits of the 
angle correlation of temperature fluctuations of the WMAP 7 year data}
\label{t:tab}
\end{center}
\end{table}

Therefore the deviation of the WMAP 7 year data angle correlations from the $\Lambda CDM$ model are within the 90.4\% confidence range corridor according 
to the Student distribution (that is in agreement with the 95\% confidence 
range \citep{BennetEtAl2011} obtained by WMAP team with the account for cosmic variance). The deviation of the WMAP 7 year data angle correlations from the equation \eqref{cfa3} model is within the narrower 81.4\% confidence range corridor according to the Student distribution, that 
is closer agreement than the $\Lambda CDM$ best fit.

The next desired step in  study of the discrete symmetries of CMBR induced by underline 
quantum gravity structure would be the measurement of triple correlations $\mean{uuu}$ and 
comparison of the resulting function with  triple correlations of galaxies density. 
The observed pattern of galaxy clustering satisfies the hierarchic form \citep{Fry1984}:
$$
\zeta_{x_1x_2x_3}=Q(\xi_{x_1x_2}\xi_{x_1x_3} + \xi_{x_1x_2}\xi_{x_2x_3}+\xi_{x_1x_3}\xi_{x_2x_3}),
$$
where $\xi_{x_i x_j}$ is the pair correlator of the observed galaxy density, with statistical estimations of the $Q$ values being in the limits $0.85\div1.24$
\citep{PG1975,GP1977}. 

To conclude with we would like to emphasize that the manifestation of quantum gravity symmetry in triple correlations 
is in some sense inverse to the measuring EPR correlations \citep{EPR1935}. In EPR the spin correlations 
are measured by {\em outside} observer, in the CMBR studies we measure correlations from {\em inside} the Universe. In 
both cases there are no {\em a priori} arguments for ergodicity, and in both cases the measurements should 
be performed {\em in situ}. The offline measurements by separate counters may just return the mean number of 
photons from each direction. 

\section*{Acknowledgements}
The authors are thankful to G. Hinshaw and D.P.Skulachev for useful discussions, and to C.Bennett 
for guidance in WMAP data sources. The authors are indebted to V.Krylov for his help in parallel computations. The research was supported in part by the RFBR Project 11-02-00604-a and the Program of Creation and Development of the National University of Science and Technology ''MISiS''.

\begin{thebibliography}{}

\bibitem[Agishtein and Migdal, 1992]{AM1992}
Agishtein, M. and Migdal, A. (1992).
\newblock {\em Modern Phys. Lett. {A}}, 7(12):1039 

\bibitem[Altaisky et~al., 1996]{ABK1996}
Altaisky, M., Bednyakov, V., and Kovalenko, S. (1996).
\newblock {\em Int. J. Theor. Phys.}, 35(2):253 

\bibitem[Ambj{\o}rn and Jurkiewicz, 1992]{AJ1992}
Ambj{\o}rn, J. and Jurkiewicz, J. (1992).
\newblock {\em Phys. Lett. B}, 278(1-2):42 

\bibitem[Barett and Crane, 1998]{BC1998}
Barett, J. and Crane, L. (1998).
\newblock {\em J. Math. Phys.}, 39:3296 

\bibitem[Bennett et~al., 1992]{Bennet1992}
Bennett, C., et al.
  (1992).
\newblock \apjl, 396:L7 

\bibitem[Bennett et~al., 1996]{Bennet1996}
Bennett, C., et al.
  (1996).
\newblock \apjl, 464:L1 

\bibitem[Bennett et~al., 2003]{Bennet2003}
Bennett, C.~L., et al. 
  (2003).
\newblock \apj, 583(1):1.

\bibitem[Boulatov, 1992]{Boulatov1992}
Boulatov, D. (1992).
\newblock {\em Modern Phys. Lett. A}, A7(18):1629 

\bibitem[Boulatov et~al., 1986]{BKKM1986}
Boulatov, D., et al.
(1986).
\newblock {\em Nuclear Physics B}, 275(4):641 

\bibitem[Br{\'e}zin and Kazakov, 1990]{BK1990b}
Br{\'e}zin, E. and Kazakov, V.~A. (1990).
\newblock {\em Phys. Lett. B}, 236(2):144 

\bibitem[{C. L. Bennett} and Wright, 2011]{BennetEtAl2011}
Bennett, C.L., et al.
(2011).
\newblock \apjs, 192(2):17.

\bibitem[Crane et~al., 2001]{CPR2001}
Crane, L., Perez, A., and Rovelli, C. (2001).
\newblock \prl, 87(18):181301

\bibitem[{D. Larson} and Wright, 2011]{Larson2011}
Larson, D., et al.
(2011).
\apjs, 192:16.

\bibitem[de~Oliveira-Costa and Tegmark, 2006]{de2006cmb}
de~Oliveira-Costa, A. and Tegmark, M. (2006).
\prd, 74(2):023005.

\bibitem[{E. Komatsu} and Wright, 2011]{Komatsu2011}
Komatsu, E., et al.
(2011).
\apjs, 192:18

\bibitem[Efstathiou, 2003]{Efstathiou2003}
Efstathiou, G. (2003).
\mnras, 343:L95 

\bibitem[Einstein et~al., 1935]{EPR1935}
Einstein, A., Podolsky, B., and Rosen, N. (1935).
\newblock {\em Phys. Rev.}, 47:777--780.

\bibitem[Fry, 1984]{Fry1984}
Fry, J. (1984).
\apj, 279(2):499 

\bibitem[Groth and Peebles, 1977]{GP1977}
Groth, E. and Peebles, P. (1977).
\apj, 217:385

\bibitem[Hamermesh, 1989]{Hamermesh}
Hamermesh, M. (1989).
\newblock {\em Group theory and its application to physical problems}.
\newblock Dover Publications.

\bibitem[Hinshaw et~al., 1996]{Hinshaw1996}
Hinshaw, G., et al.
(1996).
\apj, 464:L17

\bibitem[Hinshaw et~al., 2009]{wmap5}
Hinshaw, G.,  et~al.  (2009).
\apjs, 180:225 

\bibitem[{J. Dunkley} and Wright, 2009]{Dunkley2009}
Dunkley, J., et al.
  (2009).
\newblock \apjs, 180:306 

\bibitem[Luminet et~al., 2003]{LWRLU2003}
Luminet, J.-P., et al.
 (2003).
\nat, 425(6958):593 


\bibitem[Ooguri, 1992]{Ooguri1992}
Ooguri, H. (1992).
\newblock {\em Modern Phys. Lett.}, A7(30):2799 

\bibitem[Partridge and Wilkinson, 1967]{PW1967CMBR}
Partridge, R. and Wilkinson, D. (1967).
\prl, 18:557 

\bibitem[Peebles and Groth, 1975]{PG1975}
Peebles, P. and Groth, E. (1975).
\apj, 196:1 

\bibitem[Perez and Rovelli, 2001]{PR2001}
Perez, A. and Rovelli, C. (2001).
\prd, 63(4):041501

\bibitem[Smoot {et al.}, 1992]{Smoot1992}
Smoot, G. {et al.} (1992).
\apj, 396:L1

\bibitem[Strukov et~al., 1993]{strukov1993anisotropy}
Strukov, I., et al.
(1993).
\newblock {\em Phys. Lett. B}, 315(1-2):198 

\bibitem[Sunyaev and Zeldovich, 1970]{SZ1969s}
Sunyaev, R.~A. and Zeldovich, Y.~B. (1970).
\apss, 7(1):3 

\bibitem[Tegmark et~al., 2003]{Tegmark2003}
Tegmark, M., et al.
(2003).
\prd, 68(12):123523


\bibitem[Wright, 1992]{Wright1992}
Wright, E. (1992).
\apj, 396:L13

\end{thebibliography}

\newpage 
\appendix
\section*{Appendix}
\subsection*{Characters of irreducible representations of $C_3$ and $T_d$ groups}
\begin{table}[h]
\begin{center}
\begin{tabular}{l|lll}
 & E & $R_1$ & $R_1^2$ \\
\hline 
$ \chi^{(1)}$ & 1 & 1 & 1 \\
$ \chi^{(2)}$ & 1 & $e^\frac{2\pi\imath}{3}$ & $e^\frac{4\pi\imath}{3}$ \\
$ \chi^{(3)}$ & 1 & $e^\frac{4\pi\imath}{3}$ & $e^\frac{2\pi\imath}{3}$ 
\end{tabular}
\end{center}
\caption{Characters of irreducible representations of the $C_3$ group}
\label{t1}
\end{table}
\begin{table}[h]
\begin{center}
\begin{tabular}{l|rrrr}
 & E & $C_2(3)$ & $R_1(4)$ & $R_1^2(4)$ \\
\hline 
$ \chi^{(1)}$ & 1 & 1 & 1 & 1 \\
$ \chi^{(2)}$ & 1 & 1 & $e^\frac{2\pi\imath}{3}$ & $e^\frac{4\pi\imath}{3}$ \\
$ \chi^{(3)}$ & 1 & 1 & $e^\frac{4\pi\imath}{3}$ & $e^\frac{2\pi\imath}{3}$ \\
$ \chi^{(4)}$ & 3 &-1 & 0 & 0
\end{tabular}
\end{center}
\caption{Characters of irreducible representations of the $T_d$ group. The indices 
in parentheses denote the numbers of elements in each class}
\label{t2}
\end{table}
\end{document}